\documentstyle[dina4my,12pt,epsfig]{article}
\newcommand{\gap}{\stackrel{>}{\sim}}
\newcommand{\lap}{\stackrel{<}{\sim}}
\newcommand{\xbj}{x}
  \def\thebibliography#1{\section*{References}\list
   {[\arabic{enumi}]}{\settowidth\labelwidth{[#1]}\leftmargin\labelwidth
   \advance\leftmargin\labelsep
   \usecounter{enumi}}
   \def\newblock{\hskip .11em plus .33em minus -.07em}
   \sloppy
   \sfcode`\.=1000\relax}
  
  \newcounter{cap}
  {\begin{list}{Figure \arabic{cap}\hfil}{\usecounter{cap}
  \settowidth{\labelwidth}{Figure #1}%
  \setlength{\leftmargin}{\labelwidth}%
  \addtolength{\leftmargin}{\labelsep}%
  \setlength{\parsep}{2mm plus 1mm minus 1mm}
  \setlength{\itemsep}{3mm plus 2mm minus 2mm}
  }}%
  {\end{list}}
  {\begin{list}{}{\settowidth{\labelwidth}{#1}%
  \setlength{\leftmargin}{\labelwidth}%
  \addtolength{\leftmargin}{\labelsep}%
  \setlength{\itemsep}{0pt plus 1pt}
  \setlength{\parsep}{0pt plus 1pt}
  \setlength{\topsep}{0pt plus 1pt}
  \setlength{\partopsep}{0pt plus 1pt}
  \setlength{\parskip}{2mm plus 1mm minus 1mm}
  }}%
  {\end{list}}

\begin{document}
\input feynman
\bigphotons
\thispagestyle{empty}
\noindent
DESY 98-051                           \hfill ISSN 0418--9833 \\
     \hfill LUNFD6/(NFFL-7156) 1998
\begin{center}
  \begin{Large}
  \begin{bf}
{
Resolved photon processes in DIS and small $x$ dynamics. 
}
  \end{bf}
  \end{Large}
\end{center}
\begin{center}
  \begin{large}
H.~Jung, L.~J\"onsson, H.~K\"uster \\
    \vspace{0.5cm} 
    {\it Physics Department, Lund University, P.O. Box 118, 221 00 Lund,
Sweden}  
\end{large}
\\
\vspace*{1.cm}
  {\bf Abstract}
\end{center}
\begin{quotation}
\noindent
{\small
It has been found that recent results on forward jet production 
from deep inelastic scattering can neither be reproduced 
by models which are based on leading order $\alpha_s$ QCD matrix 
elements and parton showers
nor by next-to-leading order calculations.
The measurement of forward jet cross sections has been
suggested as a promising probe of 
new small $\xbj$ dynamics
and 
the question is
whether these data provide an indication of this.
The same question arises for
other experimental data in deep inelastic scattering at small 
$\xbj$ which 
can not be described by conventional models for deep inelastic scattering.  
In this paper the influence of resolved 
photon processes has been investigated and it has been studied to 
what extent such processes 
are able to reproduce the data.
It is shown that two DGLAP 
evolution chains from the hard scattering process towards the proton
and the photon, respectively,
 are sufficient to describe effects, observed
in the HERA data, which have been attributed to BFKL dynamics.}
\end{quotation}

\section{Introduction}
The cross section of forward jet production 
in deep inelastic scattering (DIS) 
has been
advocated as a particularly sensitive measure of 
small $\xbj$ parton dynamics
 \cite{Mueller_fjets1,Mueller_fjets2}.
Analytic calculations based on 
the BFKL equation in the  
leading logarithmic approximation (LLA)
 are in
fair agreement with data. 
However, recent calculations of the BFKL kernel in
the next-to-leading logarithmic approximation (NLLA) \cite{BFKL_NLO}
have given surprisingly large corrections, and it remains to be shown whether
the data can still be reasonably described.
\par
Monte Carlo generators 
based on direct, point-like 
photon interactions (DIR model),
 calculated from leading order (order $\alpha_s$) 
QCD matrix elements, and leading log parton showers 
based on the DGLAP evolution  do not
take 
any new parton dynamics in the small $\xbj$ region
into account and are therefore not expected
to fit the experimental data. 
Recent results from the H1~\cite{H1_fjets_data}
and ZEUS~\cite{ZEUS_fjets_data}
experiments on forward jet production 
exhibit significant deviations from the predictions of such
models. 
Also next-to-leading order calculations (NLO i.e. order $\alpha_s^2$) 
predict too small a cross section compared to data. 
\par
Similarly the DIR model is unable to reproduce
the fractional di-jet rate, $R_2$, as a function of
$\xbj$ and $Q^2$ \cite{H1_2+1jets_data},
 investigated in a kinematic domain where
$p^2_{T,jet}$
\raisebox{-.8ex} {$\stackrel{\textstyle>}{\sim}$}
$Q^2$. In these events  the two high $p_T$ jets originate from
the hard scattering process.
Discrepancies up to a factor of three are observed.
Inclusive jet cross sections in the transition region between
photo-production and DIS 
are measured to be 
significantly larger than predicted by
models assuming DGLAP based direct interactions \cite{H1_incl_jets}.
\par
Measurements of the transverse energy flow 
\cite{energyflowold,H1_eflow_prel} do not provide
a clear distinction between various models, whereas it has been
demonstrated that the data on transverse momentum spectra of 
charged particles \cite{H1_ptspectra_data} 
 can not be described by conventional 
leading order DGLAP equations at small $\xbj$ and $Q^2$, 
a region where 
 new parton dynamics are 
expected to become important.
\par
On the other hand all these data samples are rather well described by the 
color dipole model (CDM) as implemented in the Monte Carlo program
ARIADNE~\cite{CDM}. In the CDM the gluon emission is described
as radiation from color dipoles stretched between quark-antiquark pairs.
These color dipoles radiate independently and therefore the gluons are
not ordered in transverse momentum ($k_T$). 
This dis-ordering in $k_T$ is also a prominent feature of the BFKL dynamics 
where the parton emission follows a "random walk" in $k_T$.
The DGLAP evolution used in the DIR model gives strong ordering 
in $k_T$.
The question is thus whether all these data  provide
an indication of new small $\xbj$ parton dynamics.
\par
In this work we have studied the possible contribution of virtual 
resolved photons to the cross section of the DIR model as a possible 
alternative explanation for the differences between data and the 
Monte Carlo predictions. All the studies have been performed with the
RAPGAP~2.06~\cite{RAPGAP,RAPGAP206} Monte Carlo event 
generator. It has been checked
that this generator agrees well with analytical LO calculations and 
 with the LEPTO~6.5~\cite{Ingelman_LEPTO65}
Monte Carlo in the case of direct interactions.
\section{Resolved Photons in DIS}
In electron-proton scattering the internal structure of the proton as 
well as of the exchanged photon can be resolved provided the scale of 
the hard subprocess is larger than the inverse radius of the proton, 
$1/R^2_p \sim \Lambda_{QCD}^2$, and the photon, $1/R^2_{\gamma} \sim Q^2$,
respectively. Resolved photon processes play an important role in 
photo-production of high $p_T$ jets, where $Q^2 \approx 0$, 
but they can also give considerable contributions to DIS 
processes \cite{H1_incl_jets,Chyla_res_gamma} 
if the scale $\mu^2$ of the hard 
subprocess is larger than $Q^2$, the inverse size of the photon.
\par
\begin{figure}[ht]
\begin{center}
\begin{picture}(30000,25000)
\drawline\fermion[\NE\REG](5000,25000)[5000]
\drawline\fermion[\E\REG](0,25000)[5000]
\drawline\photon[\S\REG](5000,25000)[3]
\global\advance\pmidx by -5000
\put(\pmidx,\pmidy) {$y,Q^2 \to$ }
\drawline\fermion[\E\REG](\photonbackx,\photonbacky)[1500]
\global\advance\pmidy by 400
\put(\fermionbackx,\pmidy){$\bar{q}$}
\global\advance\fermionbackx by +1000
\global\advance\pbackx by 500
\global\advance\pbacky by -3200
\global\advance\Yfive by + 3800
\drawline\fermion[\S\REG](\photonbackx,\photonbacky)[1000]
\global\advance\pmidx by -4000
\drawline\gluon[\E\REG](\fermionbackx,\fermionbacky)[2]
\global\Xone = \pbackx
\global\Yone = \pbacky
\global\advance\Xone by + 1500
\global\advance\Yone by - 750
\put(\Xone,\Yone){{\Huge 
   $\searrow$} \hspace{0.5cm} $x_{\gamma}f(x_{\gamma},\mu^2,Q^2)$ 
        \hspace{0.2cm} $\gamma$ - DGLAP}
\drawline\fermion[\S\REG](\fermionbackx,\fermionbacky)[1500]

\drawline\gluon[\E\REG](\fermionbackx,\fermionbacky)[3]
\drawline\fermion[\S\REG](\fermionbackx,\fermionbacky)[2500]
\global\advance\pmidx by -5000
\put(\pmidx,\pmidy) {$x_{\gamma},\mu^2 \to $}
\drawline\fermion[\E\REG](\fermionbackx,\fermionbacky)[4500]
\global\advance\pmidy by 400
\put(\fermionbackx,\pmidy){$q$}
\global\advance\pbackx by 500
\global\advance\pbacky by -1500
\global\Xsix = \pbackx
\global\Ysix = \pbacky
\global\advance\Ysix by + 2800
\drawline\gluon[\S\REG](\fermionfrontx,\fermionfronty)[2]
\global\Xtwo = \pmidx
\global\Ytwo = \pmidy
\global\advance\Ytwo by - 500
\global\advance\Xtwo by - 4500
\put(\Xtwo,\Ytwo){{ $\hat{t} \to $}}
\global\advance\Xtwo by + 4500

\global\advance\Xtwo by + 6500
\put(\Xtwo,\Ytwo){{\Huge $\} $}}
\global\Xthree = \Xtwo
\global\advance\Xthree by + 3000
\global\advance\Ytwo by + 1000
\put(\Xthree,\Ytwo){ $qg \to qg$}
\global\Ythree = \Ytwo
\global\advance\Ythree by - 1500
\put(\Xthree,\Ythree){ hard scattering }
\drawline\gluon[\E\REG](\gluonbackx,\gluonbacky)[4]
\global\advance\pmidy by 400
\put(\gluonbackx,\pmidy){$g$}
\global\advance\pbacky by +400
\global\advance\pbacky by -1500
\drawline\gluon[\S\REG](\gluonfrontx,\gluonfronty)[3]
\global\advance\pmidx by -3000
\drawline\gluon[\E\REG](\gluonbackx,\gluonbacky)[3]
\drawline\gluon[\S\REG](\gluonfrontx,\gluonfronty)[2]
\drawline\gluon[\E\REG](\gluonbackx,\gluonbacky)[2]
\drawline\gluon[\S\REG](\gluonfrontx,\gluonfronty)[2]
\drawline\gluon[\E\REG](\gluonbackx,\gluonbacky)[1]
\global\Xfour = \pbackx
\global\Yfour = \pbacky
\global\advance\Xfour by + 2500
\global\advance\Yfour by + 2000
\put(\Xone,\Yfour){{\Huge 
   $\nearrow$}  \hspace{0.5cm}$x_{p}f(x_{p},\mu^2)$ 
                \hspace{0.2cm} $p$ - DGLAP}

\drawline\gluon[\SW\REG](\gluonfrontx,\gluonfronty)[2]

\global\advance\gluonbackx by -1500

\multiput(\gluonbackx,\gluonbacky)(0,-1000){3}{\line(1,0){9000}}
\global\advance\gluonbacky by -1000
\global\advance\gluonbackx by -500
\put(\gluonbackx,\gluonbacky){\oval(1000,3000)}
\global\advance\gluonbackx by +2000
\global\advance\gluonbacky by +1000
\global\advance\gluonbackx by -2000
\global\advance\gluonbacky by -1000
\global\advance\gluonbackx by -500
\drawline\fermion[\W\REG](\gluonbackx,\gluonbacky)[2000]
\global\advance\fermionbacky by -3000
\global\advance\fermionbackx by 4000
\global\advance\fermionbacky by 3000
\global\advance\fermionbackx by -4000
\global\advance\pmidy by 500
\put(\pbackx,\pmidy){$p$}
\end{picture}
\end{center}
\caption{Deep inelastic scattering with a resolved virtual photon and
the $q_{\gamma} g_p \to q g $ partonic subprocess.
\label{resgam1} }
\end{figure}
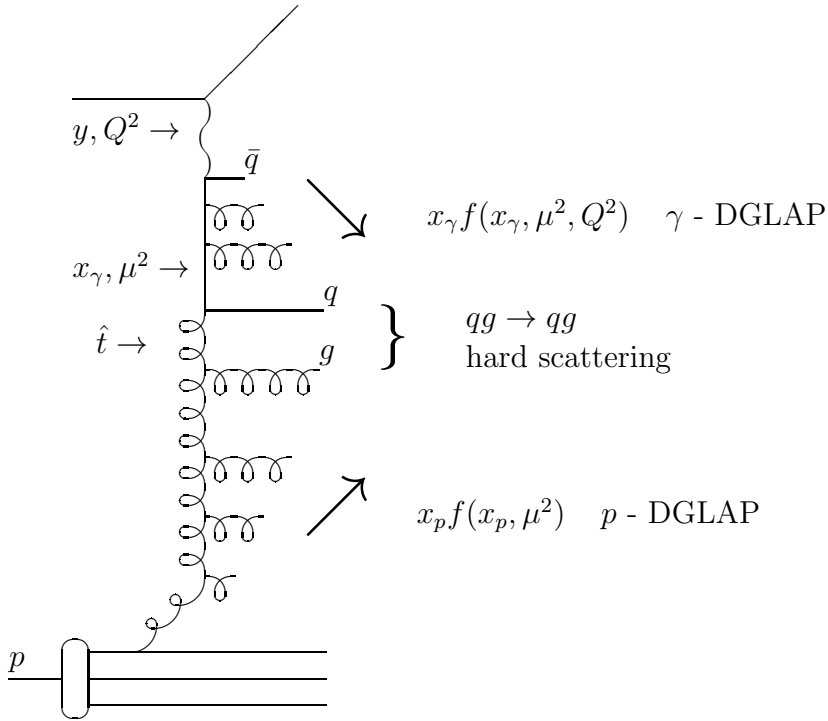

In the following we give a brief description of the model for resolved virtual
photons used in the Monte Carlo generator RAPGAP.
Given the fractional momentum transfer of the incoming electron to the 
exchanged photon, the Equivalent Photon Approximation provides the 
flux of virtual transversely polarized photons 
\cite[and references therein]{RAPGAP206,RAPGAP}.
The contribution from longitudinally polarized photons has been neglected.
The partonic structure of the virtual photon is defined by 
parameterizations of the parton densities,
$x_{\gamma} f_{\gamma}(x_{\gamma},\mu^2,Q^2)$, which  
depend on the two scales $\mu^2$ and $Q^2$
\cite{GRS,Sasgam,Drees_Godbole}. 
The following hard subprocesses are considered (RES model): 
$ gg \rightarrow q \bar{q}$,
$ g g \rightarrow gg$, 
$ q g \rightarrow q g $, 
$ q \bar{q} \rightarrow g g $, 
$ q \bar{q} \rightarrow q \bar{q}$,
$ q q \rightarrow q q $. 
Parton showers on both the proton and the photon side are included.
The generic diagram for the process $ q_{\gamma} g_{p} \rightarrow q g $
including parton showers is shown in Fig.~\ref{resgam1}.

Since the photon structure function depends on the scale, $\mu^2$, of the 
hard scattering process, the cross section of resolved photon processes 
will consequently also depend on the choice of this scale. 
The parton density of the photon is evolved from a starting scale $Q^2_0$
 to the scale $\mu^2$, the virtuality 
at the hard subprocess, giving a resummation to all orders.
\par
It should be noted that a NLO calculation assuming point-like virtual
photons contains a significant part of what is
attributed to the resolved structure of the virtual photon in the 
RES model~\cite{Kramer_Poetter_dijets}.
This is due to the fact that in order  $\alpha_s^2$ the virtual photon
can split into a $q \bar{q}$ pair with one of the two quarks interacting
with a parton from the proton, giving rise to two high $p_T$ jets.

\subsection{Choice of Scale}
In leading order $\alpha_s$ processes
the renormalization scale $\mu_R$  
and factorization scale $\mu_F$ are not well defined
which allows a number of reasonable choices.
There are essentially two competing effects:
a large scale suppresses $\alpha_s(\mu^2)$ but
gives, on the other hand, an increased
parton density, $xf(x,\mu^2)$, for a fixed small $x$ value.
The net effect depends on the details of the interaction 
and on the parton density parameterization.
\par
However, in resolved virtual photon processes the choice of
the scale $\mu^2$, at which the photon is probed, is severely
restricted \cite{gosta_torbjorn}. If we proceed from the electron
(see Fig.~\ref{resgam1}) towards the hard subprocess, the first
branching occurs already  at the electron vertex $e \to e' \gamma^*$, 
where the scale in DIS is chosen to be $Q^2=-t$
($t$ being the Mandelstam variable of a $2 \to 2$ process),
 which is related 
to the transverse momentum of the scattered electron by:
$Q^2 = p_{T;e'}^2/(1 - y)$. Here $y$ is defined by $y=\frac{q.P}{l.P}$, with
$l$ and $P$ being the four vectors of the incoming electron and proton and 
$q$ is the four vector of the (virtual) photon with $q^2=-Q^2$.
This indicates that the scale must be larger than $p_{T;e'} ^2$,
for a typical value of $y\sim 0.5$, this gives $Q^2 = 2 \cdot p_{T;e'}^2$
\par
Within a consistent picture of the evolution from the electron towards the
hard scattering process, the definition of the scale should be kept the same
for all branchings. Thus the proper scale for a 
 hard scattering process in the $\hat{t}$ - channel (as shown in 
Fig.~\ref{resgam1}) would be $\hat{t}$, since $Q^2 = -t$.
Here $\hat{t}$ refers to the hard scattering process.
 However not only
 $\hat{t}$ - channel
diagrams contribute, but also $\hat{u}$ and $\hat{s}$ - channel diagrams
 as well as
interference terms.
The calculation of the relative contributions to the cross section in 
principally limited due to the interference terms. 
We are therefore forced to make a best guess of a reasonable scale. 
We have chosen 2 different scales:
$\mu^2 = 4 p_T^2$ and $\mu^2 = Q^2 + p_T^2$. The first 
one is
closely related to the $\hat{t}$ - channel diagrams,
where we have included a factor of $4$ to account for the fact that 
$|\hat{t}| >p_T^2$.
The second one is chosen for a smooth transition
 from usual DIS to
DIS including resolved photons and to photo-production. 
This choice of scale 
has also been used in
NLO calculations including resolved photons in deep inelastic scattering 
\cite{Kramer_Poetter_dijets}.

\subsection{Parton Distribution Functions}
Two parameterizations of the parton distribution in the proton, 
GRV 94 HO (DIS) and CTEQ4M have been considered, 
which both give good agreement with
the proton structure function
 data \cite{H1_F2,ZEUS_F2}. It was found that
the produced results were identical within the percent level,
when keeping $\Lambda_{QCD}$ fixed.  
\par
Also for the virtual photon two different parameterizations of the
parton density have been investigated.
The photon can interact via its
partons either in a bound vector meson state or as decoupled partons if 
the $p_T$ of the partons is high enough. 
Only the latter is relevant in the $Q^2$ range considered in this paper.
The splitting
$\gamma \to q \bar{q}$  is called 
the anomalous component of the photon. The SaS~\cite{Sasgam} parameterization 
offers a choice of $Q_0^2$ values at which the anomalous part becomes effective. 
We have studied 
these choices resulting in different magnitudes of the
parton densities, and consequently of the cross sections. 
For the SaS 
parameterization  we have used  $Q_0^2$ as given by eq.(12) of ref.~\cite{Sasgam}
 ($IP2 = 2$), which gave the largest resolved photon contribution.
\par
 The GRV LO parton density description of
the real photon together with the 
virtual photon suppression factor of Drees and Godbole (DG) \cite{Drees_Godbole}
gives similar results.
Drees and Godbole proposed a $x_{\gamma}$ - independent suppression factor:
$$ r = 1 - \frac{\log (1 - \frac{Q^2}{P_c ^2})}
                {\log (1 - \frac{\mu ^2}{P_c^2})}$$
with an adjustable parameter $P_c^2$, which is set to 
$P_c^2 =0.5$ GeV$^2$ typical for a hadronic scale.
 Comparisons between the SaS and the DG 
parameterizations are given for some of the observables.
\par
The GRS \cite{GRS} structure function for the virtual photon has not been used, 
since it is restricted to $Q^2 < 10 $ GeV$^2$ which is not useful for the 
studies presented here,
\par
If not stated otherwise, in the following we have used 
the scale $\mu^2=Q^2+p_T^2$,
the CTEQ4M parameterization of the parton density in the proton and
the DG description of the photon.

\section{Forward Jets}
HERA has extended the available $\xbj$ region down to values 
below $10^{-4}$ where new parton dynamics 
might show up.
Based on calculations in the LLA of the BFKL kernel, the cross section for
DIS events
at low $\xbj$ and large $Q^2$ with a high $p^2_T$ jet in the 
proton direction (a forward jet) \cite{Mueller_fjets1,Mueller_fjets2} is
expected to rise more rapidly with decreasing $\xbj$ than expected
from DGLAP based calculations.
New preliminary results from the H1~\cite{H1_fjets_data} and 
ZEUS~\cite{ZEUS_fjets_data}
experiments have recently been 
presented.
The data can be described neither by conventional DIR Monte Carlo models
nor by a NLO calculation, while
comparisons to analytic calculations of the
LLA BFKL
mechanism has proven reasonable agreement.
\par
It should be kept in mind that both the 
NLO calculations and the BFKL based
calculations are performed on the parton level whereas the 
data are on the level of hadrons. 
Nevertheless these comparisons make sense, considering the present
measuring errors and the fact that simulations have shown that the
hadronization effects are at most 20\%.
\par
In this study, 
Monte Carlo events from the RAPGAP generator have been used  
to investigate whether the experimental data can be equally well  
reproduced by the inclusion of resolved photon processes.
The cuts applied in the laboratory frame are equivalent to those in
the preliminary H1 analysis \cite{H1_fjets_data}.

\noindent
Kinematic cuts: 
\begin{eqnarray}
y& > &0.1 \\ 
E'_e &> & 11 \mbox{GeV} 
\end{eqnarray}
Jet selection: 
\begin{eqnarray}
E_{jet} & >  &28.7  \mbox{ GeV } \\
p_T^{jet} & > & 3.5 \mbox{ GeV } \\
7^o & < \theta _{jet} < &20^o
\end{eqnarray}
where $\theta = 0^o$ corresponds to the proton beam direction.
\par
Jets were reconstructed in the 
$\eta$-$\phi$ space
by applying a cone algorithm 
with a cone radius of $R = \sqrt{\eta^2 + \phi^2} = 1$.
The jet cuts have been defined to ensure that
the momentum fraction
of the jet ($x_{jet} = E_{jet}/E_p$, where $E_p$ is the proton energy) 
 is large compared to $\xbj$
so as to maximize the phase space for 
a evolution in $\xbj$.
At large $x_{jet}$ the parton distributions of the proton are well
measured and the uncertainty from non-perturbative effects is 
avoided. 
In order to suppress the phase space
for DGLAP evolution, 
the $p^2_T$ of the jets  was further 
required to be of the same order of magnitude as $Q^2$ ($0.5 < p^2_T/Q^2 < 2$).
Thus DIS events with small $\xbj$ and an energetic high $p_T$ jet 
in the forward direction are selected.

\begin{figure}[htb]
\begin{center}
\epsfig{figure=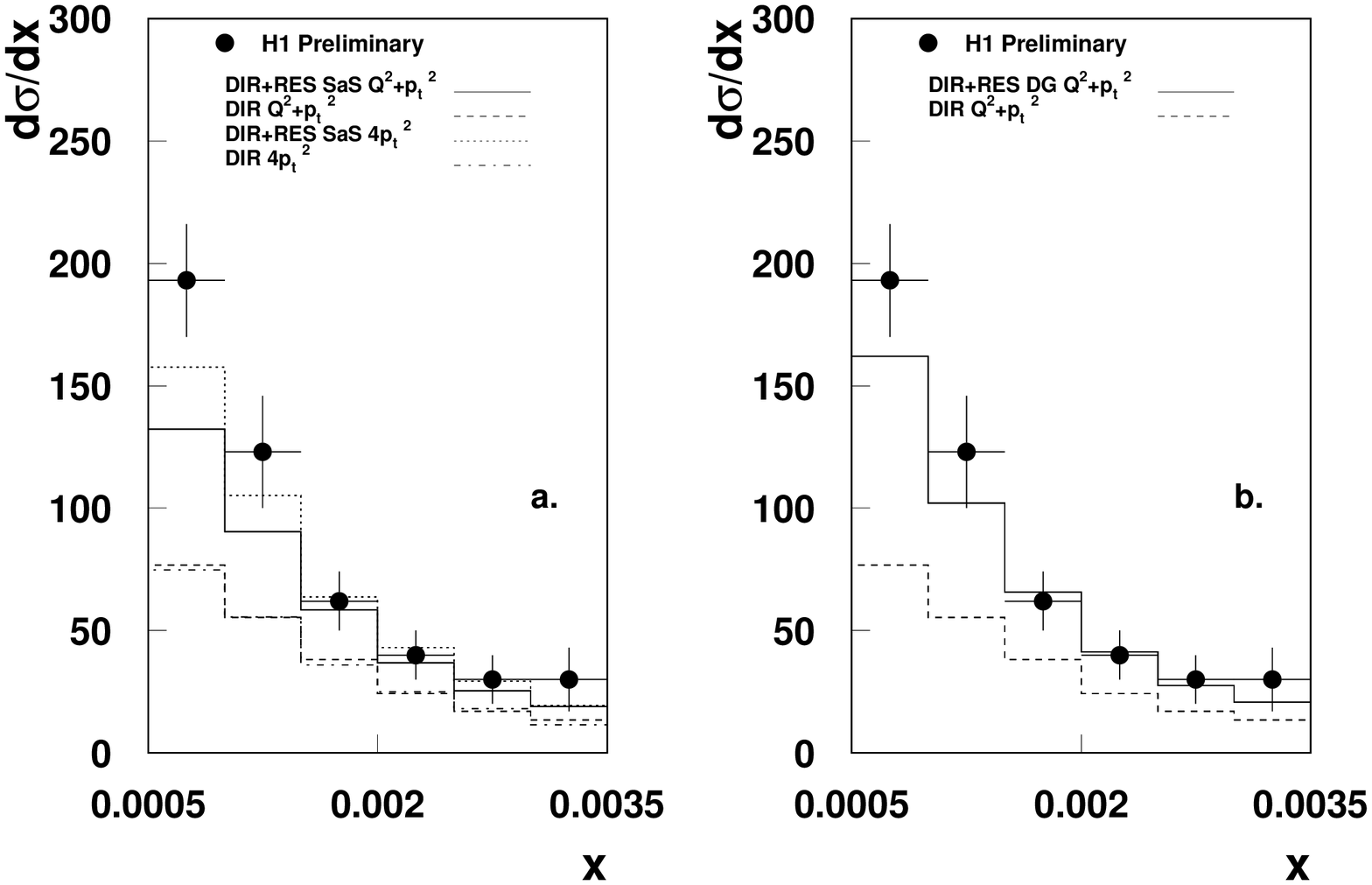,
width=18cm,height=12cm}
\end{center}
\caption{
The forward jet cross section as a function of $\xbj$ using the
cuts specified in the text. The data points
are preliminary H1 data \protect\cite{H1_fjets_data}. 
In $a.$ are shown
the RAPGAP predictions 
for the sum 
of direct and resolved processes using again the scales $\mu^2=Q^2+p_T^2$ 
(solid line) and $\mu^2=4p_T^2$ (dotted line).
Also included are the predictions for direct 
DIS processes with the scales $\mu^2=Q^2+p_T^2$ (dashed line) and 
$\mu^2=4p_T^2$ (dash-dotted line), respectively.  
The parton distribution of the virtual photon was parameterized according to
SaS 2 DIS. In $b.$  the sum of direct and 
resolved processes (solid line) and the direct (dashed line) are 
presented using the DG + GRV description 
of the parton density in the photon and with the scale $\mu^2=Q^2+p_T^2$. 
In all cases the parton density of the proton was described by CTEQ4M.
\label{fjet_res+dir}}
\end{figure}

In Fig.~\ref{fjet_res+dir} the preliminary data of the 
H1 collaboration  \cite{H1_fjets_data} are 
compared to the RAPGAP Monte Carlo predictions for both, 
direct process alone (labeled DIR) and for
the sum of direct and resolved processes (labeled DIR+RES).
The comparison is performed for the two 
scales, $\mu^2=Q^2+p_T^2$ and $\mu^2=4p_T^2$ and for the two 
parameterizations of the photon parton density 
(SaS, Fig.~\ref{fjet_res+dir}a, and DG, Fig.~\ref{fjet_res+dir}b). 
It is clearly seen that
the direct contribution alone is not sufficient to reproduce the
data whereas the DIR+RES model gives
good agreement with data.
The scale $\mu^2=4p_T^2$ seems to give a slightly better agreement 
with data than the scale $\mu^2=Q^2+p_T^2$
for the SaS photon structure function whereas for the DG photon structure
function the scale $\mu^2=Q^2+p_T^2$ leads to a good description of data
and $\mu^2=4p_T^2$ results in predictions which overshoot the data
(not shown). 
This simply illustrates the uncertainty 
in the scale and in the parameterization of the photon structure function.
The hard process $q_{\gamma}g_p \to qg$ 
contributes most to the forward jet cross section.

\begin{figure}[htb]
\begin{center}
\epsfig{figure=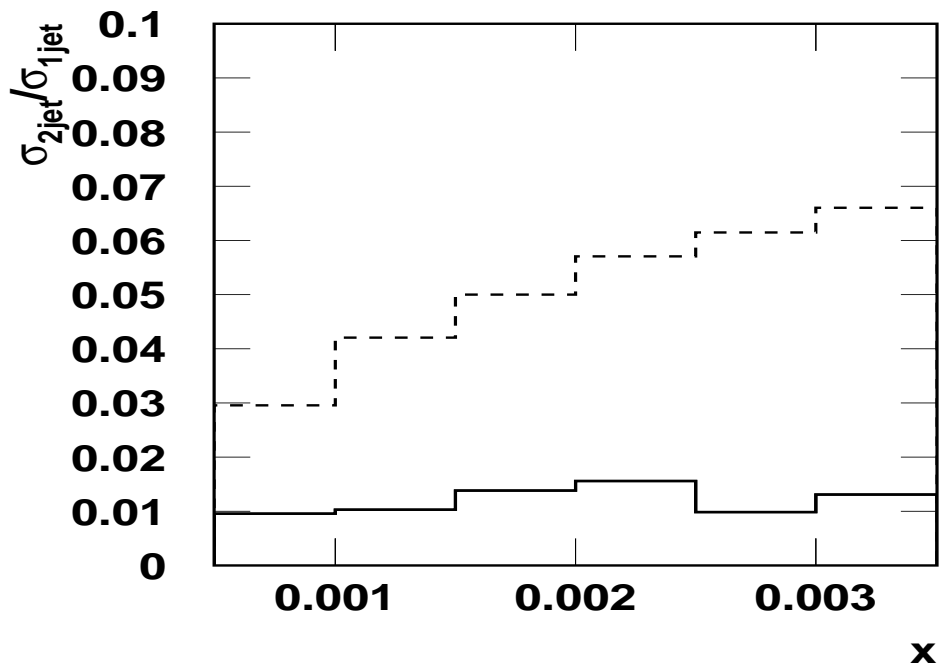,
width=16cm,height=12cm}
\end{center}
\caption{
The ratio of cross sections for the production of two and
one forward jets versus $x$, as predicted by the RAPGAP Monte Carlo
for the sum of direct and resolved processes (solid line).
 The cuts applied are 
specified in the text. The prediction from 
the analytical calculation of \protect\cite{BFKL_dijets}
is shown with the dashed line.
\label{fjet_dijets}}
\end{figure}

A small fraction of the DIS events, fulfilling the selection criteria
for forward jets, actually contains two identified jets. The ratio of
the cross sections for events 
with two and one forward jet(s) would be a measure of the BFKL
vertex function which controls the emission of the second jet. 
Analytic calculations (in LLA) \cite{BFKL_dijets}
have been performed in the same kinematic
region and with the same jet selection as defined above for the one-jet sample.
The predicted ratio varies from $3 \%$ to $6 \%$, for a cone radius of $R=1$,
when going from $\xbj = 0.5 \cdot 10^{-3}$ to $\xbj = 3 \cdot 10^{-3}$. 
Comparisons to experimental data are not yet available.
The prediction of the RAPGAP Monte Carlo for direct plus resolved
processes is shown in Fig.~\ref{fjet_dijets}.
It is about a factor of 3 lower than the prediction from the 
BFKL calculations.
A large part of this discrepancy could be due to hadronization effects
which would reduce the prediction of the parton level BFKL calculation.
Given the large uncertainties both in the resolved photon 
and BFKL calculations,  
it is not obvious that the two approaches  give results in the same order of
magnitude.

\section{Other tests of low $\xbj$ dynamics}
In this section we discuss measurements like the $(2+1)$ jet rate in DIS,
transverse energy flow and the $p_T$ spectra of charged particles,
which are not well described by 
DIR models and have been subject to speculations on their possible
interpretation in terms of small $\xbj$ BFKL dynamics.
\subsection{Di-jet rates in DIS}
New H1 data on fractional di-jet rates  
were recently made 
public \cite{H1_2+1jets_data}. 
Events with two jets were selected in the kinematic region
$5 \lap Q^2 \lap 100 $ GeV$^2$ and $10^{-4} \lap \xbj \lap 10^{-2}$. 
The jets were defined in the hadronic center-of-mass system  
using a cone algorithm of radius $R=1$ and requiring a transverse
momentum $p_T^{jet}> 5$ GeV. The results make evident
that standard DIR LO Monte Carlo models such as 
RAPGAP~2.06 in the direct mode and 
LEPTO~6.5 badly fail to describe the data.
Especially in the $Q^2$ region below 30 GeV$^2$ the discrepancies are
large but in this region $p_T^2 > Q^2$ and therefore resolved photon 
interactions are expected to play a significant role. The CDM, on the 
other hand, gives very good agreement with data.
\par 
With the same cuts as in the H1 analysis, RAPGAP 
has been used, with and without resolved photon processes included, to
predict the di-jet rates. 

\begin{figure}[htb]
\begin{center}
\epsfig{figure=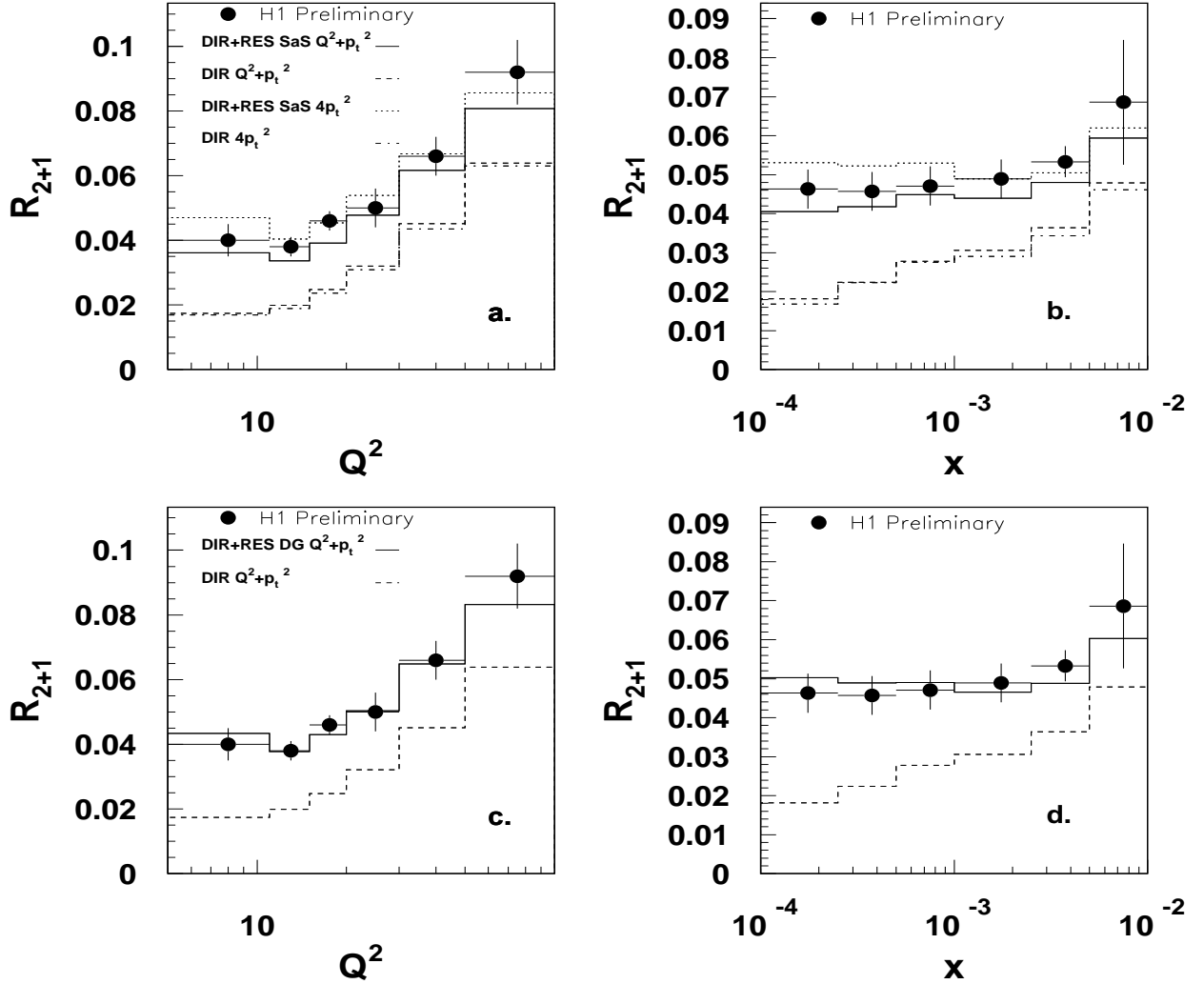,width=18cm,height=16cm}
\end{center}
\caption{The di - jet ratio $R_{2+1}$ as a function of
$Q^2$ ($a.$ and $c.$) and $\xbj$ ($b.$ and $d.$). 
The points   
are preliminary H1 data \protect\cite{H1_2+1jets_data}. 
In $a.$  and $b.$ are shown the RAPGAP predictions 
 for the sum 
of direct and resolved processes using the scales 
$\mu^2=Q^2+p_T^2$ 
(solid line) and $\mu^2=4p_T^2$ (dotted line),
as well as the predictions 
for direct 
processes with the scales $\mu^2=Q^2+p_T^2$ (dashed line) and 
$\mu^2=4p_T^2$ (dashed-dotted line). 
The SaS 2 DIS parameterization was used to describe the parton density 
in the virtual photon. In $c.$ and $d.$ 
 the 
sum of direct and resolved processes (solid line) 
and the direct (dashed line)  
are presented using 
the DG description of the parton density in the photon  with the 
scale $\mu^2=Q^2+p_T^2$. The parton density of 
the proton is given in all cases by the CTEQ4M parameterization.
\label{2+1jet_res+dir}} 
\end{figure}

The data from H1 are shown as a function of $Q^2$ and $\xbj$ in 
Figs. ~\ref{2+1jet_res+dir} a) and b), respectively, 
together with the predictions from 
RAPGAP. 
 The RAPGAP results
on direct and direct+resolved processes are shown separately 
for the two scales mentioned above and for the 
two parameterizations of the photon structure. The data are well above the
direct contribution, but
adding the direct and resolved  contributions gives a
remarkably good description of data all the way down to the phase 
space region where the resolved processes are dominating,  
 especially with the scale 
$\mu^2=4 p_T^2$ ($\mu^2=Q^2+p_T^2$) and the SaS (DG) parameterization
 of the photon structure.
Also in this data sample the  major contribution is coming from the
$q_{\gamma}g_p \to qg$ subprocess.

\subsection{Inclusive jet production}

\begin{figure}[htb]  
\begin{center}
\epsfig{figure=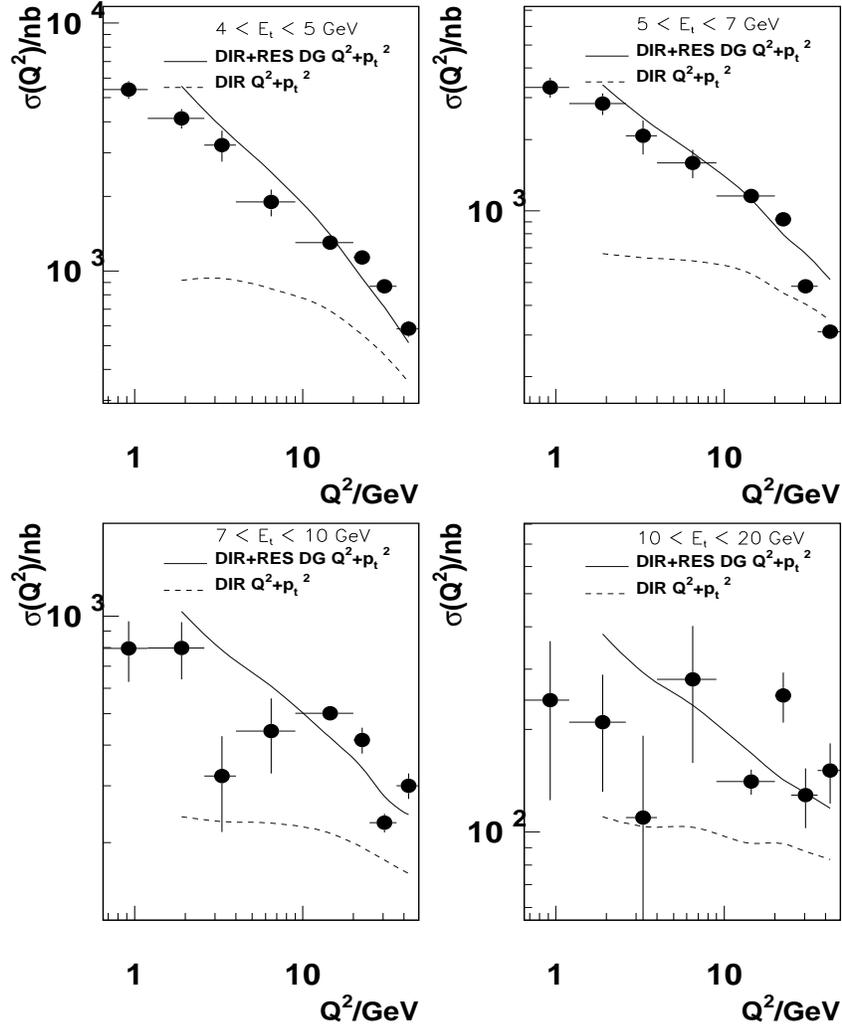,width=14cm,height=16cm}
\end{center}
\caption{The inclusive jet cross section as a function of
$Q^2$ for different regions of the jet $E_T$. 
The  points are published H1 data~\protect\cite{H1_incl_jets}, 
  the curves represent
the RAPGAP predictions 
for the sum of direct and resolved processes (solid line) and 
for direct processes only (dashed line).
\label{incl_jet}}
\end{figure}

The H1 experiment has measured the single inclusive jet cross section in
a range of photon virtualities from $Q^2$ = 0 to 50 GeV$^2$ using the
$k_T$ cluster algorithm in the hadronic CMS~\cite{H1_incl_jets}.
 The jets from DIS processes were 
required to have transverse momenta above 4 GeV/c.
 Fig.~\ref{incl_jet}  shows the inclusive jet cross section as
a function of $Q^2$ in bins of $E_T$ together with the predictions of 
the DIR model and DIR+RES model.
 In the region where 
$Q^2 \gap E_T^2$ the DIR model is expected to be valid since the photon 
can not be resolved. It is  found that the DIR model approaches 
the data as $Q^2$ increases whereas in the low $Q^2$ region the deviations
become significant. The DIR+RES model gives reasonable agreement with data
in the full kinematic range investigated.

\subsection{Transverse energy flow}
\begin{figure}[htb]
\begin{center}
\epsfig{figure=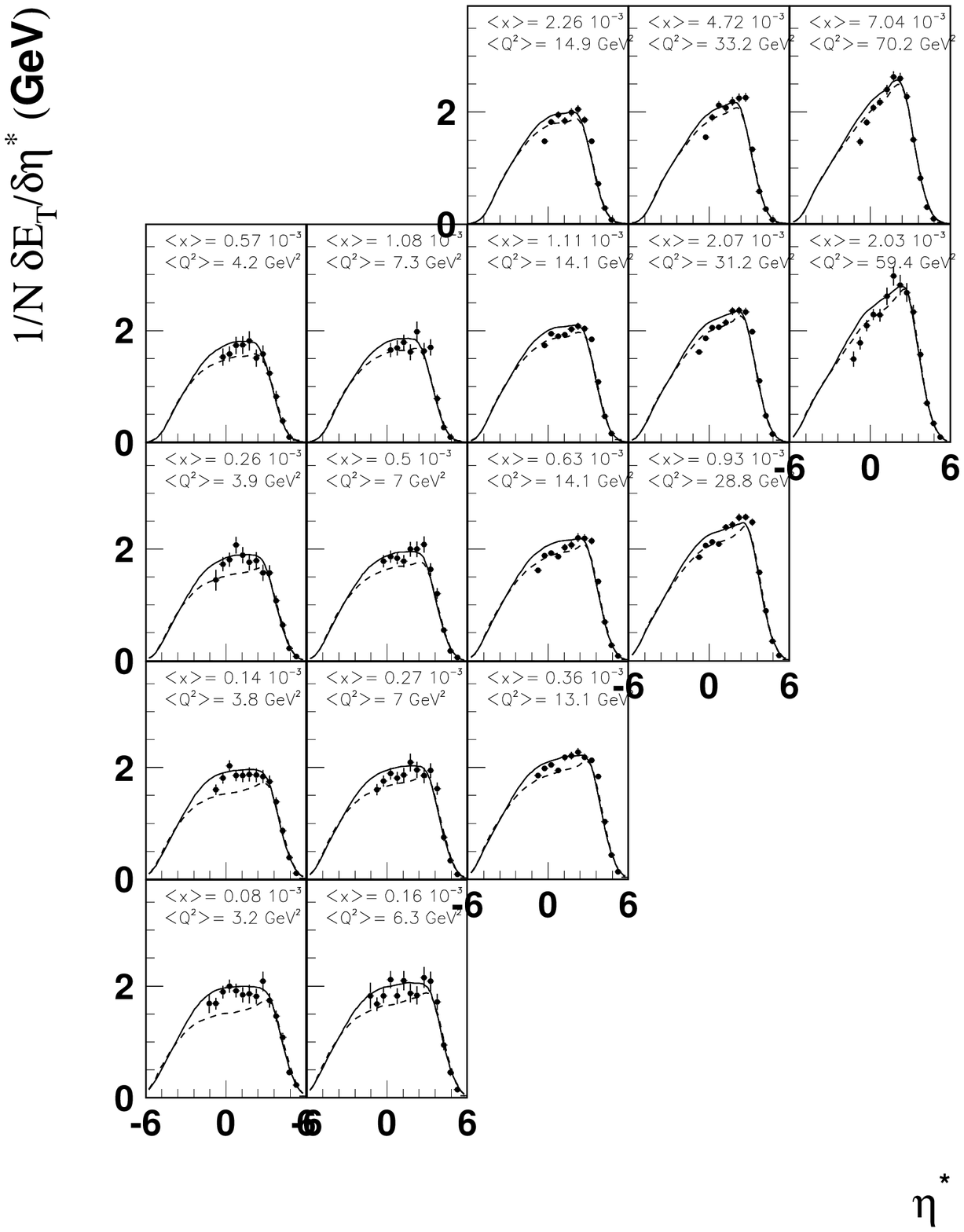,width=18cm,height=18cm}
\end{center}
\caption{The transverse energy flow as a function of $\eta^*$ in the
hadronic CMS for different bins in $Q^2$ and $\xbj$. 
The data points are preliminary H1 results~\protect\cite{H1_eflow_prel}.
Negative $\eta^*$ corresponds to the proton fragmentation region.
The 
curves give the RAPGAP predictions for 
 the sum of direct and resolved processes (solid line) and
for the direct contribution only (dashed line).
\label{energy_flow}}
\end{figure}

 The non ordered $k_T$ parton emissions in low $\xbj$ BFKL dynamics
 produce more transverse energy than the standard DGLAP evolution scheme 
in the central rapidity region in the hadronic CMS, which corresponds to the
forward region in the laboratory system ($\theta \approx 10^o$).
 Results from H1~\cite{H1_eflow_prel} on the transverse energy flow,
 presented in the hadronic CMS as a function of the 
 pseudo-rapidity, $\eta^*$, are shown in Fig.~\ref{energy_flow}
 for the kinematic region 3 GeV$^2 < Q^2 < 70$ GeV$^2$ and
 $8 \cdot 10^{-5} < \xbj < 7 \cdot 10^{-3}$.
\par
The largest difference between LLA BFKL and standard DGLAP calculations are 
 expected at small $\xbj$ and central rapidities $\eta^*$.
Discrepancies between data and the DIR model of
 refs.~\cite{energyflowold,MEPS} 
are also observed in this kinematic region.
 Only after including a proper treatment of
 the parton dynamics
 of the proton remnant in the sea quark scattering processes
 \cite{Ingelman_LEPTO65}
 and using a rather low
 cutoff on the gluon emission for the parton cascade, a reasonable 
 agreement is achieved with the exception of the
 lowest $\xbj$ and $Q^2$ region (see Fig.~\ref{energy_flow}).  
However the description is significantly improved by including
the contribution of resolved virtual photons giving 
  an excellent description of the   
 transverse energy flow  over the full range of
 $\xbj$ and $Q^2$ as illustrated in Fig.~\ref{energy_flow}.
Different scales $\mu^2$ or the parton distribution functions give changes of
the order of $5 \%$.
\subsection{$p_T$ spectra of charged particles}
Studies based on QCD models have demonstrated that the high $p_T$
tail of 
 charged particle transverse momentum spectra
 is sensitive to small $\xbj$ dynamics of parton
radiation and that the influence from hadronization is small
\cite{Kuhlena,Kuhlenb}.
 The LLA BFKL dynamics result in  harder $p_T$ spectra 
than obtained in the DGLAP scenario
because of the non-ordered $k_T$ emissions.
A LLA BFKL based calculation~\cite{BFKL_ptspectra}
including 
an estimation of the fragmentation effects was able to reproduce
the high $p_T$ tail in the lowest $\xbj$ bins. 
\begin{figure}[htb]
\begin{center}
\epsfig{figure=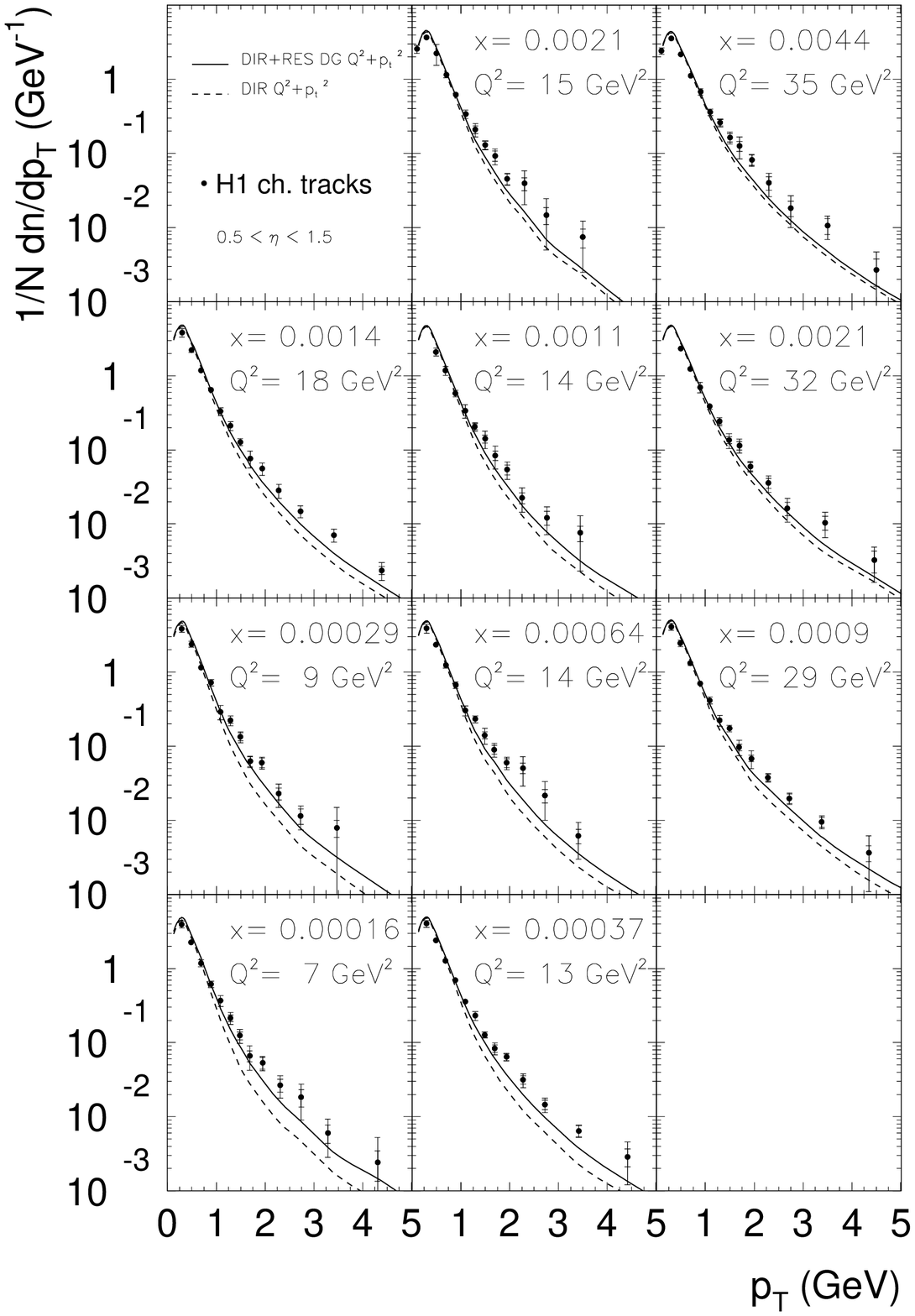,width=15cm,height=18cm}
\end{center}
\caption{Transverse momentum distributions of charged particles
  in the rapidity range $0.5 < \eta^* < 1.5 $ in the 
hadronic CMS in bins of $Q^2$ and $x$ as obtained by 
H1~\protect\cite{H1_ptspectra_data}.
The RAPGAP prediction for direct processes (dashed line)
and for the sum of direct and resolved photon contributions (solid line) 
are shown for comparison.
\label{pt_spectra}}
\end{figure}

Fig.~\ref{pt_spectra} shows the $p_T$ distributions of charged
particles as measured by the H1 collaboration~\cite{H1_ptspectra_data}
for DIS events with 
3 GeV$^2 < Q^2 < 70$ GeV$^2$ in the rapidity range
$0.5 < \eta < 1.5$.
The  H1 analysis \cite{H1_ptspectra_data}
showed that both, direct models and
CDM provide good agreement with the  data at large $\xbj$, whereas
with decreasing $\xbj$ the direct models start falling below the data.
 In Fig.~\ref{pt_spectra} the dashed line shows the
prediction of the DIR model and the solid line the 
prediction of the DIR+RES model as calculated 
with the RAPGAP Monte Carlo program. The solid line in 
Fig.~\ref{pt_spectra} demonstrates a good description of the data
over the full range in $\xbj$ and $Q^2$.

\section{Discussion}

Recent experimental data on forward jet production show deviations from
traditional LO Monte Carlo models assuming  
directly interacting point-like photons. This has given
 rise to various speculations.
Similar deviations
have been observed in other measurements like
the fractional di-jet rates, inclusive jet production, 
transverse energy flow and in the transverse momentum spectra of 
charged particles.
It is tempting to assume that the observed effects could be explained by
BFKL dynamics 
as firstly the 
 data on forward jet production, which has been proposed 
as a possible probe of low $\xbj$ dynamics, can not be reproduced by the 
DGLAP evolution and
secondly  
 that the CDM MC, which produces
radiation without ordering in $k_T$ as expected for BFKL, reproduces all data
fairly well.  

In the present study it has been shown that the addition of resolved 
photon processes to the direct interactions in DIS
leads to good agreement with the data, 
without invoking any BFKL parton dynamics.
We have observed that the dominant contributions to the resolved photon
processes come from order $\alpha_s^2$ diagrams with the hard 
subprocess $q_{\gamma} g_p \to q g $ (see Fig.~\ref{resgam1}).
 Since the partons which form
the photon remnant per definition have smaller $p_T$ than the 
partons involved in the hard scattering, a situation with non $k_T$
ordering is created.
\par
In the LO DIR model the ladder of gluon emissions is 
governed by DGLAP dynamics giving a strong ordering of 
$k_t$ for emissions between the photon and the proton vertex.
The models describing resolved photon processes and BFKL dynamics 
are similar in the sense that both lead to a breaking of this
ordering in $k_t$.
The BFKL picture, however, allows for complete dis-ordering in
$k_t$, while in the resolved photon case the DGLAP ladder is split into
two shorter ladders,
one from the hard subsystem to the proton vertex, and one to the photon
vertex, each of them ordered in $k_t$ (see Fig.~\ref{resgam1}).
 Only if the ladders are long enough to produce 
additional hard radiation it might be possible to separate resolved 
photon processes from processes governed by BFKL dynamics.
Thus the resolved photon approach may be a ``sufficiently good''
approximation to an exact BFKL calculation and the two approaches may prove 
indistinguishable within the range of $\xbj$ accessible at HERA. 
\par
It should be emphasized again that a NLO calculation 
assuming point-like virtual
photons contains a significant part of what is
attributed to the resolved structure of the virtual photon in the 
RES model~\cite{Kramer_Poetter_dijets}.
\section{Acknowledgments}

It is a pleasure to thank G. Ingelman and A. Edin for discussions
about the concept of resolved photons. We have
also profited from a continuous dialogue with B. Andersson, G. Gustafson
and T. Sj\"ostrand. We want to thank G. Kramer and B. P\"otter for many
discussions on the relation between
 resolved photons in DIS and its relation to NLO
calculations. 
We are grateful to J. Gayler for careful reading of the
manuscript and for clarifying comments. 
We also want to thank M. W\"usthoff and J. Bartels
 for discussions on BFKL and the resolved photons in DIS.

\end{document}